# Non-line-of-sight reconstruction via structure sparsity regularization

**DUOLAN HUANG,[1] QUAN CHEN,[1] ZHUN WEI,[2] RUI CHEN[1,*]**

[1]*School of Physics & State Key Laboratory of Optoelectronics Materials and Technologies, Sun Yat-sen University, Guangzhou 510275, China*
[2]*College of Information Science and Electronic Engineering, Zhejiang University, Hangzhou 310027, China*
*\*Corresponding author: chenr229@mail.sysu.edu.cn*

**Non-line-of-sight (NLOS) imaging allows for the imaging of objects around a corner, which enables potential applications in various fields such as autonomous driving, robotic vision, medical imaging, security monitoring, etc. However, the quality of reconstruction is challenged by low signal-noise-ratio (SNR) measurements. In this study, we present a regularization method, referred to as structure sparsity (SS) regularization, for denoising in NLOS reconstruction. By exploiting the prior knowledge of structure sparseness, we incorporate nuclear norm penalization into the cost function of directional light-cone transform (DLCT) model for NLOS imaging system. This incorporation effectively integrates the neighborhood information associated with the directional albedo, thereby facilitating the denoising process. Subsequently, the reconstruction is achieved by optimizing a directional albedo model with SS regularization using fast iterative shrinkage-thresholding algorithm. Notably, the robust reconstruction of occluded objects is observed. Through comprehensive evaluations conducted on both synthetic and experimental datasets, we demonstrate that the proposed approach yields high-quality reconstructions, surpassing the state-of-the-art reconstruction algorithms, especially in scenarios involving short exposure and low SNR measurements.**

Non-line-of-sight (NLOS) imaging is a computational imaging technique that reconstructs the object around the corner by capturing and analyzing the multiple scattered light emanating from the hidden object. This technique has become a rapidly growing field due to its potential applications in fields such as autonomous driving, robotic vision, medical imaging, security monitoring, etc. Among various imaging approaches, transient NLOS imaging stands out for its ability to generate three-dimensional reconstruction by exploiting the rich temporal information obtained from a time-resolved sensor [1-9]. Specially, in a transient system, a light pulse that illuminates a visible wall is redirected to the hidden scene which scatters the photons back to the wall, and then a time-resolved single-photon detector toward the wall captures the back-scattered signal. By illuminating and focusing at different positions of the wall, a series of time-resolved signal is obtained, namely transient measurements [1]. These measurements encode the information about hidden scenes, including the object's spatial position, shape, albedo and surface normal. Thus, NLOS imaging can be mathematically formulated as an inverse problem to reconstruct the hidden scene's information from transient measurements.

A fundamental challenge for high-quality NLOS reconstruction is the low signal-noise ratio (SNR) of transient measurements. The radiometric falloff of the multiple scattered photons results in low-flux signal photons that hit the sensor, which induces both Gaussian noise from the background and Poisson noise in the single-photon detection [10]. The systematic methods to improve SNR include increasing laser power and the exposure time to increase the number of photons that reach the sensor, but these methods are often demanding for real applications [8]. Algorithmic approaches rely on accurate imaging models and robust inversion algorithms. Wave-based models retrieve the hidden scenes by wave propagation, which is robust to reflectance property and multiple scattering [8, 11]. However, these methods only reconstruct the albedo, without modeling the surface normal. Time-of-flight based models describe the light transport by transient rendering, which maps the scene's albedo distribution and surface normal to transient measurements [2,12]. The transient rendering model is accurate yet highly nonlinear, thus most methods employ the linearized version that parameterizes the hidden scene as a 3D albedo volume that absorbs the surface normal information [1,4]. Several other methods investigate temporal discontinuity to directly invert the rendering model [5,13]. Filtered back-projection (FBP) algorithm provides a non-iterative method to derive an approximate solution to the linearized model [1, 4,14]. The introduction of the light-cone transform (LCT) in confocal NLOS system allows further simplification of NLOS reconstruction as a deconvolution problem, enabling fast closed-form solution and low memory cost [15]. Directional LCT (DLCT) method extended the LCT to allow for the simultaneous reconstruction of the hidden scene's albedo and surface normal, and it also alleviate noise by Wiener filtering [6].

However, Wiener filtering assumes the presence of only white noise whereas the measurements are corrupted by the mixture of Gaussian and Poisson noise. As a result, it is not desirable for NLOS denoising, leading to low-contrast reconstructions, artefacts and difficulties in normal estimation. To tackle the ill-posedness and denoising problem, the incorporation of regularization is a typical method that leverages scene or system priors, such as total variations (TV) [15,16] and curvature [17] regularizations. The iterative LCT method achieves this by regularizing solution with TV and $l_1$-norm, which encourages piecewise smoothness and albedo sparseness, respectively [15]. However, TV has the drawbacks of oversmoothing in low-contrast region and generating the staircase effect. In response to these challenges, Liu et al. proposed a signal-object collaborative regularization (SOCR) approach by incorporating albedo sparseness, non-local similarity and signal smoothness [18] and further extended it to accommodate arbitrary illumination and detection pattern without dense measurements [19]. This joint regularization of the hidden object and transient measurements yields high-quality reconstruction. However, it is noted that the forward model of SOCR necessitates substantial matrix storage, and the optimization process becomes intricate due to the extensive number of hyperparameters involved [18,19]. Consequently, there is an ongoing need to reduce the complexity of the optimization process while preserving reconstruction quality.

Compared to TV that integrates a penalty of variation completely localized, structure tensor TV (STV) offers a more robust measure of local variation by considering variational information around a local neighborhood, leading to superior performance in image denoising [20,21]. These works involving STV inspire us to leverage the directional-albedo information within a local neighborhood to enhance measurements of the hidden object's structure details and preserve it from noise. Here, we present a novel regularization called structure sparsity (SS) for denoising in NLOS albedo-normal joint reconstruction. Instead of assuming albedo sparseness [18], we assume structure sparseness to the hidden scene. To be specific, a voxel-wise matrix is created by stacking the directional-albedos within a local window as the column vectors to account for neighborhood information, and the structure is represented by its singular values and vectors. Then, the nuclear norm is penalized to enforce structure sparsity. Its effectiveness is validated on both synthetic and experimental datasets. Results indicate that the approach offers superior reconstruction quality compared to other regularization methods when reconstructing with short exposure and low SNR measurements.

In this paper, the schematic diagram of NLOS imaging process as shown in Fig.1 is formulated by the forward directional albedo model [6], the details of which can be found in Supplement 1. Based on the discretized model in Eq.(S2) of supplement 1, the overall optimization problem is formulated as:

$$\boldsymbol{\rho}^* = \arg\min_{\rho}\left[\frac{1}{2}\|H(\boldsymbol{\rho})-\tau\|_2^2 + \lambda\|\|W\boldsymbol{\rho}\|_*\|_1\right], \qquad (1)$$

where $\|\cdot\|_2$ represents the Euclidean distance, $\boldsymbol{\rho}$ denotes the directional albedo volume and $H(\cdot)$ denotes the discrete directional albedo model. The first term is the data fidelity term measuring the closeness of the estimated transient intensity $H(\boldsymbol{\rho})$ to the transient measurement $\boldsymbol{\tau}$. Assuming the domain of interest is discretized into $N^3$ voxels, the directional albedo vector at the $n^{th}$ voxel, $\boldsymbol{\rho}_n = \rho_n \boldsymbol{n}_n$ represents the hidden scene's albedo $\rho$ and surface normal $\boldsymbol{n}$ at the $n^{th}$ voxel, respectively.

The second term corresponds to the proposed SS regularization term, designed to enhance denoising in NLOS reconstruction. The key idea involves eliminating noise from the NLOS structure by leveraging the structure sparsity prior. To achieve this, the structure information at the $n^{th}$ voxel of object domain (the center voxel in Fig.1(b)) can be extracted from the directional-albedo information within a local window (the red square in Fig.1). Typically, this local window is composed of voxels with the $n^{th}$ voxel at the center, forming a local cubic region with $L$ voxels, and a 2D illustration of which is shown in Fig.1(b). Then, a voxel-wise matrix $(W\boldsymbol{\rho})_n$ of size $3\times L$ is created by storing the weighted directional albedo within the local window of the $n^{th}$ voxel:

$$(W\boldsymbol{\rho})_n = \left[ \sqrt{w_1} \begin{pmatrix} \rho_x \\ \rho_y \\ \rho_z \end{pmatrix}_{R_n(1)}, \sqrt{w_2} \begin{pmatrix} \rho_x \\ \rho_y \\ \rho_z \end{pmatrix}_{R_n(2)}, \cdots, \sqrt{w_L} \begin{pmatrix} \rho_x \\ \rho_y \\ \rho_z \end{pmatrix}_{R_n(L)} \right], \qquad (2)$$

where $R_n(l)$ denotes the index of the $l^{th}$ neighborhood voxel within the local window of the $n^{th}$ voxel, and $w_l$ ($l = 1, 2,..., L$) denotes the weight assigned to the $l^{th}$ neighborhood voxel, which follows a Gaussian distribution with variation $\sigma$ (see Eq.(S5) of Supplement 1). When the total number of the voxel in the local window, $L$, is set to 1, implying the exclusion of the neighborhood information, we refer to it as "local SS". This scenario represents a special case of the proposed SS regularization (See Section 2 of Supplement 1).

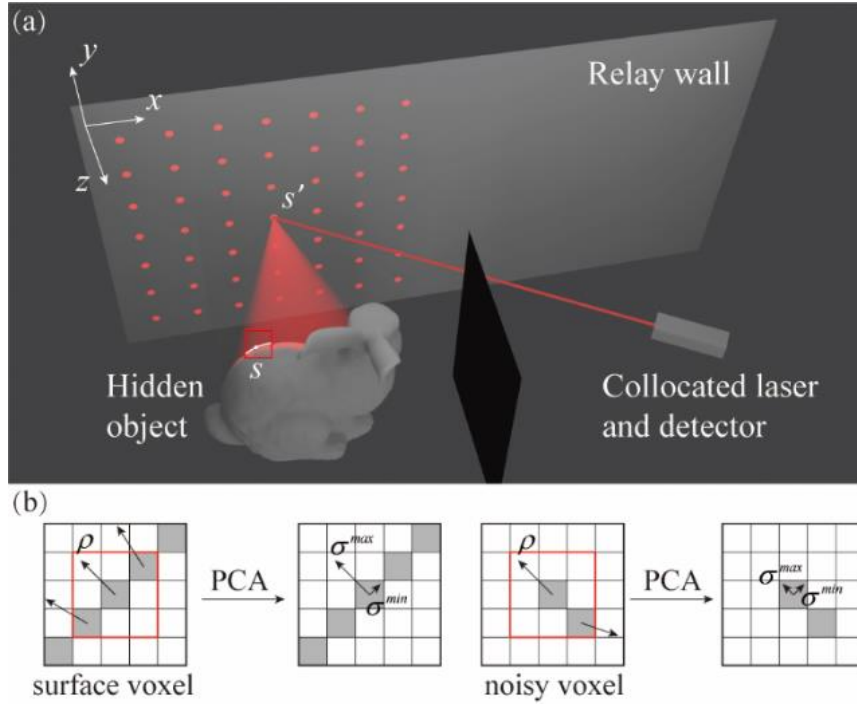

Fig. 1. Transient NLOS imaging with SS regularization. (a) The schematic diagram of NLOS imaging process and (b) SS regularization analysis with the surface voxel and noise voxel.

Conducting singular value decomposition (SVD) on this matrix of $(W\boldsymbol{\rho})_n$ is equivalent to identifying the principal directions that best characterize the normal and albedo information at the $n^{th}$ voxel, similar to principal component analysis (PCA), as depicted in Fig.1(b). After SVD (see Eq. (S6) of Supplement 1), we obtain three singular vectors of $\boldsymbol{u}_n^{max}$, $\boldsymbol{u}_n^{mid}$ and $\boldsymbol{u}_n^{min}$, which denote the primary directions voted by the neighborhood directional albedos. The corresponding singular values denote the voted albedo intensity along these directions. Intuitively, as depicted in Fig.1(b), a voxel located on the hidden surface tends to have a large singular value $\sigma_n^{max}$ and two relatively small ones of $\sigma_n^{mid}$ and $\sigma_n^{min}$, whereas for a noisy outlier voxel all singular values tend to be very small, thus penalizing the sparsity of the singular values, namely SS regularization, helps to remove noise from clean hidden structures.

To enforce SS, we penalize the $l_1$-norm of singular values of $W\boldsymbol{\rho}$, equivalent to the nuclear norm of $W\boldsymbol{\rho}$, denoted by $\|W\boldsymbol{\rho}\|_*$. As a result, the overall regularization term becomes the $l_1$-norm of the nuclear norm $\left\| \|W\boldsymbol{\rho}\|_* \right\|_1$. The hyperparameter $\lambda$ denotes the trade-off between data fidelity and scene prior, and its determination can follow the strategy outlined in [22]. Since the SS regularization term is not differentiable everywhere, the optimization problem cannot be solved by gradient-based methods. We employ the FISTA for its ability to minimize the combination of a Lipschitz-continuous function and a non-differentiable function with the state-of-the-art convergence rate. The computational complexity of the proposed approach is analyzed and the algorithm flow is also provided (See Supplement 1).

To demonstrate the effectiveness of the proposed method, the reconstruction performance is evaluated on both synthetic datasets and experimental datasets. We compare the proposed SS method with other state-of-the-art methods including F-K migration (FK) [8], phasor field (PF) [11], SOCR [18], and different regularizations attached to the same directional albedo model, including Wiener filtering (DLCT method) [6], $l_1$ sparsity regularization, and local SS. $L$ in the local window is fixed to 27, and the kernel is Gaussian with variation $\sigma$ set to 0.5 for synthetic data and 0.35 for real data. Note that we downsample the signals to reduce the computation time by averaging the signals within a temporal or spatial window, which improves the SNR in preprocessing step.

The Zaragoza NLOS synthetic dataset is employed to qualitatively and quantitatively evaluate the performance of the proposed method [23]. For the first example of planar scene "T", its synthetic data 0.5 m from the wall is measured on a 0.6 m x 0.6 m wall with scanning resolution of 64 x 64 and temporal resolution of 0.0025 m (assuming c = 1m/s). The data includes photons with multiple scattering. Figure 2 shows the reconstruction scene of "T" using both noiseless data (the first row) and noisy data (the second row) for different methods. For the noiseless data, it is found that DLCT, FK and PF reconstruction in Fig. 2(a), 2(c) and 2(d) is blurred by artefacts caused by the low-pass property of the system, while other approaches in Fig. 2(b) and 2 (e-g) are able to reduce artefacts and recover clear structure of the letter "T".

For the noisy case, the noisy data is generated by simulating Poisson and Gaussian noise on the noiseless data (see Section 4 in Supplement 1). As shown in Fig. 2(h-n), only SOCR and SS methods preserve clear structures while other methods fail to suppress the noise. To be clear, the quantitative indicators including structure similarity index measure (SSIM) and peak signal-to-noise ratio (PSNR) that compare the reconstructed results with the ground truth are also provided (See Table S1 of Supplement 1), from which the similar phenomena can be obviously attained.

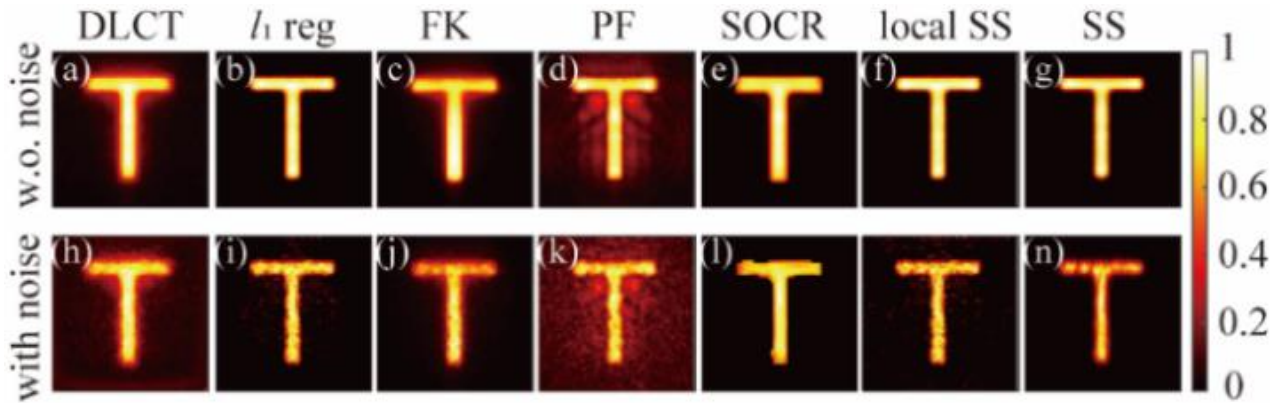

Fig. 2. Reconstruction of planar synthetic scene "T" for different method.

The second synthetic example of 3D scene "bunny" 0.5m from the wall was collected on a 1m x 1m wall with scanning resolution of 256 x 256 and temporal resolution is 0.003m (assuming c = 1m/s). We downsample the spatial resolution to 64 x 64 and the temporal resolution to 0.006m. Figure 3 shows the reconstructions for the different regularization methods. Compared to the planar scene, the three components of surface normal are separately provided for the 3D scene. It is showed that DLCT reconstructions in Fig. 3(a-c) contain noise and artefacts due to the intrinsic ill-conditioned property of the inverse problem. $l_1$ regularization in Fig.3(d-f) produces a sparser result compared to DLCT method, however, it tends to oversmooth the $x$ and $y$ components while retaining noisy points in the $z$ component due to the absence of consideration of directional correlation. For the SOCR method displayed in Fig.3(g-i), it yields high-contrast reconstruction, however, some detailed structure such as the bunny's ear is lost. In contrast, both local SS and SS regularization in Fig.3(j-o) provide clear structures with few artefacts due to the exploitation of the correlation between directional components. SS regularization also results in shaper edges and more accurate depth recovery (See Table S3 of Supplement 1). Furthermore, a comparison of the albedo volume of the "bunny" scene using various methods as scene "T" is also provided (see Fig.S1 of Supplement 1).

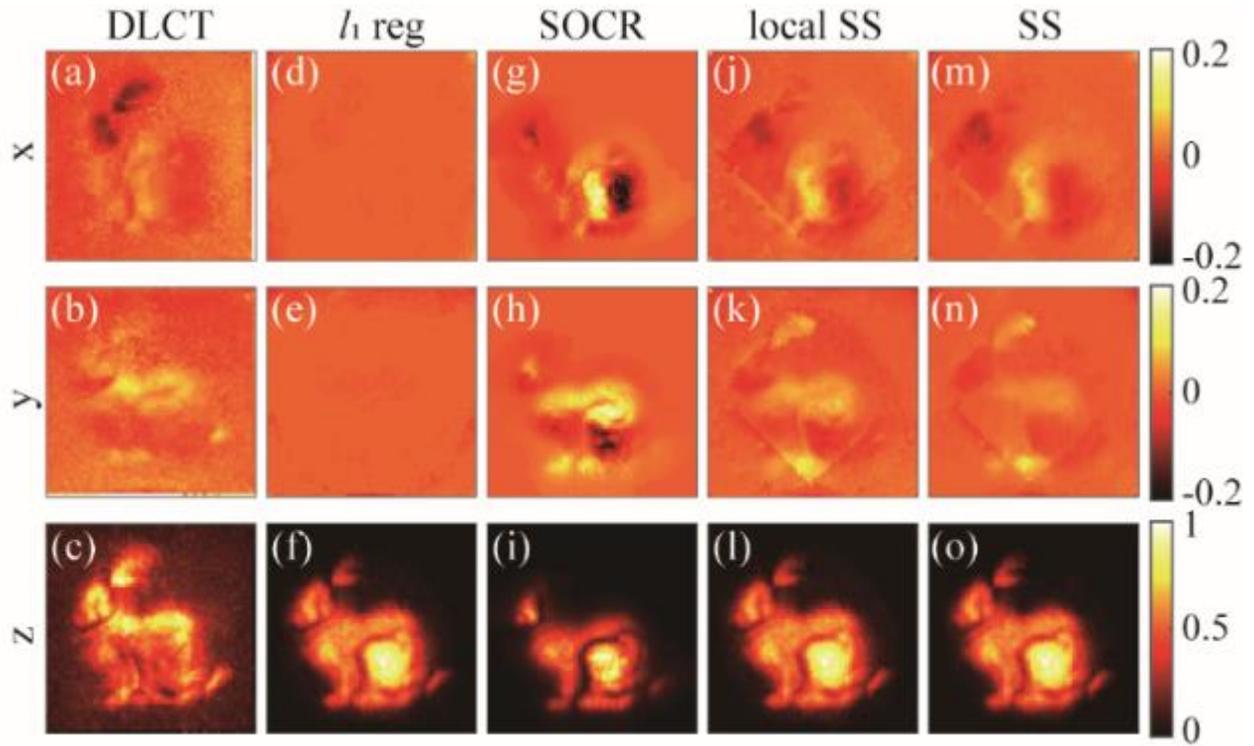

Fig. 3. Reconstructed directional albedo of synthetic scene "bunny" using different methods

Further, Stanford experimental datasets are used for validation of the proposed SS regularization method [8,15]. We reconstruct the measurements of resolution chart, dot chart and a mannequin. All measurements are conducted on a 0.4m x 0.4m wall with scanning resolution of 64 x 64 and temporal resolution of 4ps. We downsample the temporal resolution to 16ps. For these cases in Fig. 4, DLCT, FK and PF methods produce noisy image background with low contrast due to their limited denoising ability, as depicted in the first, third and fourth columns of Fig.4. $l_1$ regularization, local SS and SOCR could mitigate background noises, but the detailed structure is slightly corrupted, which is reflected in the second, fifth and sixth columns of Fig.4. SS regularization (the seventh column of Fig.4) can substantially remove noise and recover the detailed structure due to the consideration of neighborhood information. For clarity, the cross sections for different methods are compared (See Fig. S2 of Supplement 1), which confirms the ability of the proposed method to recover fine structure from noisy measurements.

To evaluate denoising performances of the proposed method, we reconstruct the "dragon" scene from experimental data with different noise levels. The exposure time is 30min and 1min, corresponding to measurements with high SNR and low SNR, respectively. The scanning area is 2m x 2m with spatial resolution of 512 x 512 and temporal resolution of 32ps. We downsample the spatial resolution to 128 x 128 and the temporal resolution to 64ps. The reconstruction of high-SNR measurements (i.e. 30-minute exposure in Fig.5(a)) shows that $l_1$ regularization, SOCR, local SS and SS regularizations can generate sharper images than DLCT. $l_1$ regularization tends to oversmooth the $x$ and $y$ components. SS regularization produces a slightly shaper results compared with other methods. In the case of low-SNR reconstruction of Fig.5(b), the DLCT method fails to suppress noise. $l_1$ regularization can only partially suppress the noise in $z$ direction component even when $x$ and $y$ component are oversmoothed. Local SS could not remove the noisy image part that is close to the detailed structure. SOCR provides decent images in both noise levels, however the reconstruction is slightly degraded in low SNR scenario. Among the above algorithms, SS regularization (the last row of Fig. 5) achieves best reconstruction quality, which implies that the proposed method is robust towards noise (See Fig. S3 of Supplement 1 for 15 second exposure time reconstruction). Note that even in low-SNR system, SS regularization is still able to provide result that is comparative to that of high-SNR system, which indicates that the proposed method has the potential in help reducing the exposure time in real applications.

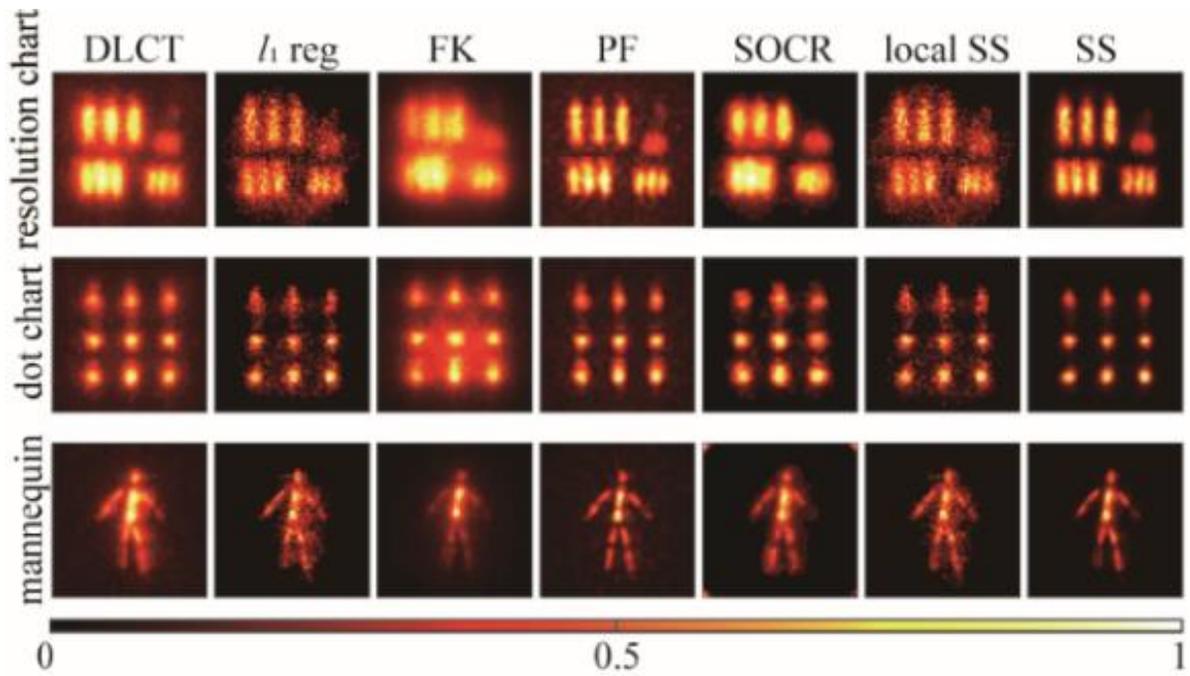

Fig. 4. Reconstruction of experimental scenes of "resolution chart", "dot chart" and a mannequin using different methods.

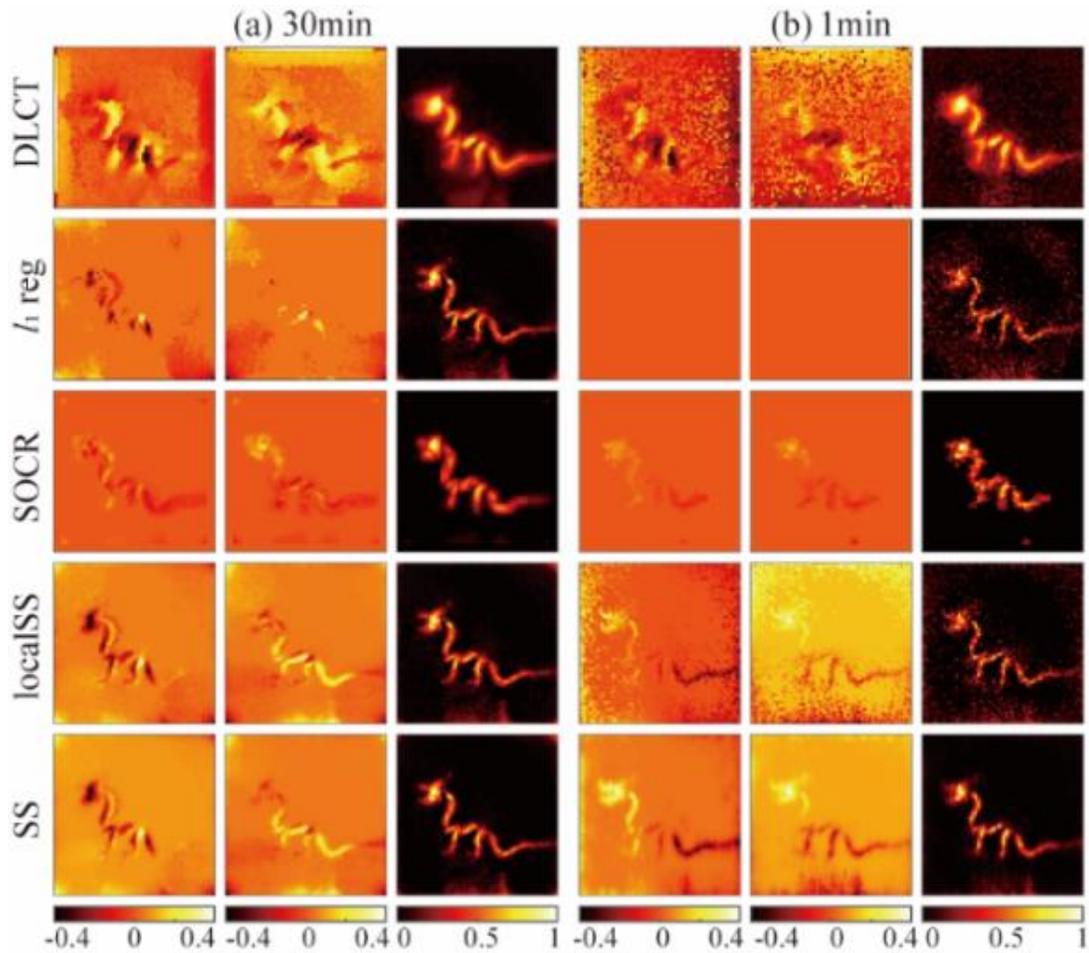

Fig. 5. Reconstruction of experimental scene "dragon" using the measurements with the exposure time (a) 30 min and (b) 1min.

In conclusion, we present a novel regularization method known as SS regularization for denoising and high-quality albedo-normal reconstruction in NLOS imaging. The reconstruction is achieved through the optimization of a regularized directional albedo imaging model. The proposed SS regularization displays superior performance in denoising and preserving structure details when compared to other reconstruction methods, which has the potential to facilitate high-quality reconstruction in scenarios with a short exposure time. We should note that the SVD operation results in high computational cost and long processing time. Besides, we only consider the case of confocal setting, however, we believe that this method could be transformed to nonconfocal settings by simply employing a nonconfocal directional albedo imaging model.

**Funding.** Guangdong Basic and Applied Basic Research Foundation (2023B1515040023); Guangzhou Science and Technology Program (202201011671); State Key Laboratory of Extreme Photonics and Instrumentation

**Disclosures.** The authors declare no conflicts of interest.

**Data availability.** Data underlying the results presented in this paper are not publicly available at this time but may be obtained from the authors upon reasonable request.

**Supplemental document.** See Supplement 1 for supporting content.